# Structure-Aware Multimodal Retrieval Under Data Scarcity, Privacy Constraints and Out-of-Domain Content

Argyrios Zafeiriou[a,*], Dionysios Koulouris[a], Ilias Maglogiannis[a]

[a]Department of Digital Systems, University of Piraeus, 80 Karaoli and Dimitriou Street, 18534 Piraeus, Greece
[*]Corresponding author. E-mail address: zafeiriou@unipi.gr (A. Zafeiriou)

## Abstract

**Purpose** Figure retrieval from reference atlases in expert domains like forensic pathology, dentistry and geology must work without labels or training pairs, on content that often cannot leave local hardware, and on images outside the training data of image-text encoders. This work proposes a structure-aware pipeline that satisfies all three, retrieving figures through the page structure recovered by layout geometry and a local open-weight 8B VLM.
**Methods** A geometry-based pairing method is introduced, in which the model receives a schematic of the detected boxes and selects each picture's caption without access to pixels or text. Caption retrieval is evaluated on three private corpora (29 books, 10,421 figures, 218 queries) with blind human grading, against a lexical baseline, generated descriptions and five image-text encoders.
**Results** Geometric pairing was correct for 96%, 92% and 77% of pictures in the three domains, against 76%, 45% and 53% for the better of two document parsers. A hosted model given the schematic was equally accurate, while coordinates as text lost 22 points. Caption retrieval obtained an nDCG@5 of 0.85, 0.87 and 0.76, against 0.54, 0.51 and 0.27 for the best cross-modal encoder and 0.41, 0.28 and 0.30 for generated descriptions. Pairing errors carried over to retrieval.
**Conclusion** The proposed pipeline achieves accurate figure retrieval with small local models and no in-domain training, where general and biomedical image-text encoders fall far behind. Its geometry-only pairing is the most accurate of five methods and exposes no content, even to a hosted model. Code and prompts are released[1].

## 1. Introduction

Reference atlases hold rich visual knowledge of expert domains such as forensic pathology, oral medicine and petrography, and practitioners practically consult atlases for visual demonstration of relevant findings. The books are distributed as large PDF files with dense and irregular layouts, in which a page may carry several pictures with captions placed below, beside or above them. The unit that answers a retrieval query is therefore the figure together with its caption, whereas recent document retrievers operate on whole pages [1–3]. Often, collections of this kind are the sole data sources for expert domains, and without external annotated and multimodal datasets available, fine-tuning or selecting retrieval models is not feasible. Furthermore, the academic content is mostly licensed, not allowing redistribution and open-sourcing of foundation models, and in the case of medicolegal data, it is also sensitive, excluding hosted services to be used and requiring private model instances to operate on local, resource limited hardware. Generic image-text encoders [4, 5], though they provide strong data representations for retrieval purposes, are trained on web photographs, biomedical ones on curated collections of medical images [6, 7], and both can provide narrower expertise than that met within the mixed contents of specialized atlases. The available methods perform well on the documents and images they were developed for, but each of them depends on something that these collections do not offer. Matching the query directly to pixels needs an encoder that covers the domain, and describing every figure with a vision-language model (VLM) only moves this dependence to the describing model, which must also be small enough to run locally. Document parsers extract figures and captions [8, 9], but existing tools are known to misidentify figures and captions when the structure of the page is not trivial [10], which is the normal case in atlases.

[1] https://github.com/zafeiriou-arg/structure-aware-figure-retrieval.

This work starts from the hypothesis that successful and generalizable parsing of the structure of the documents, especially that of figure-caption pairs, can substitute for the lack of sufficient training corpora and in-domain foundation models, and the usage of external non-private services. The proposed pipeline parses each page with a layout detector, pairs every picture with its caption, and indexes the resulting figures with a multilingual text encoder, using only open-weight models that run locally and without any training. For the pairing step a geometry-only method is introduced, in which a small VLM is shown a schematic of the detected boxes and asked which box is the caption of a given picture. The decision thus rests on the arrangement of the page, and the model never sees its pixels or its text.

The pipeline is evaluated on three private corpora from forensic pathology, dental and oral medicine and geology, comprising 29 books, 10,421 figures and 218 retrieval queries whose results were graded blind to the retrieval method. Comparators for the benchmarks include generic and medical foundation models, and document parsing algorithms. Retrieval through the paired captions reaches an nDCG@5 between 0.76 and 0.87, against 0.27 to 0.54 for the best of five general and biomedical image-text encoders and 0.28 to 0.41 for descriptions generated by the same local model. The geometric pairing is correct for 77% to 96% of the pictures, against 45% to 76% for the better of two established document parsers, and exceeds a nearest-caption rule by up to 31 points. The contributions of this work are the following.

- A geometry-only method for figure-caption pairing, in which a small VLM decides from a drawn schematic of the detected boxes. The same model is 22 points less accurate when the boxes are given as coordinates in text, and a hosted model given the schematic is as accurate as the local one, so the step can be delegated without exposing any content.
- A comparison of five pairing methods on atlas pages with several pictures. The geometric pairing is the most accurate overall and at least as accurate as each of the others in every domain examined, while two established document parsers pair only 44% and 57% of the pictures with their correct caption.
- Evidence that pairing errors carry over to retrieval. With captions linked by a document parser, the first result shows the queried finding for 45% of the dental queries, against 96% with the geometric pairing.
- An evaluation with human-graded relevance in three expert domains, showing that the recovered captions outperform generated descriptions and five general and biomedical image-text encoders, and that BM25 on the same captions is already better than every route that does not use them.
- An evaluation protocol for private corpora without labels, in which the queries are drafted from the topic names and the pooled results are graded blind. The code and the prompts of the pipeline are released for applications to similarly behaving corpora.

Section 2 reviews related work, Section 3 describes the pipeline, the data and the evaluation protocol, Section 4 reports the results, Section 5 discusses them together with the limitations of the study, and Section 6 concludes.

# 2. Related Work

## 2.1 Document parsing and figure-caption pairing

Figure extraction from PDF files has been studied mainly on scholarly papers, with PDFFigures 2.0 extracting figures together with their captions [11], DeepFigures learning figure detection from distantly supervised data [12], and PDFigCapX being motivated by the observation that existing parsing tools often misidentify figures and captions when the structure of the page is not trivial [10]. Furthermore, layout detectors trained on DocLayNet [13] and open toolkits such as Docling [8] and MinerU [9] have made page parsing widely available, while recent parsers rely on VLMs for the same purpose [14, 15]. Although these tools perform well on the documents they were developed for, a comparison of ten parsing tools over six document categories reported that the best tool depends on the document type, with all of them struggling on scientific and patent documents [16], while benchmarks such as OmniDocBench cover an increasingly diverse range of documents [17]. Reference atlases lie at the edge of this range, since a page often carries several pictures with their captions on different sides, and, to our knowledge, the reliability of figure-caption pairing on such pages has not been evaluated. The present work treats this pairing as a separate step, proposing a method that requires only the geometry of the detected boxes and comparing its accuracy against a nearest-caption rule, a VLM that reads the page, and the Docling and MinerU parsers.

## 2.2 Retrieval and image-text encoders in specialized domains

Visual document retrievers, embedding page images with VLMs, have become the reference approach for visually rich documents [1–3], supported by benchmarks of growing size and difficulty [18, 19], and, recently, have also been applied to geological documents, which we also include in this study [20]. However, in the case of material centered on figures, text representations were recently found to be the most effective in scientific retrieval, even for queries about figures, since they carry the captions and the surrounding context [21]. Similarly, descriptions generated by a large VLM and embedded with a text encoder surpassed specialized visual document encoders without any contrastive training [22], while figure-caption pairs mined from open-access papers form the basis of retrieval benchmarks and medical datasets [23–25]. Although these findings support retrieval through text, they were obtained on public documents with the figure-caption pairs already available and, in the case of generated descriptions, with large models. Regarding the direct matching of text and images, generic encoders are trained on image-text pairs from the web [4, 5] and biomedical ones on figure-caption pairs from the research literature or on curated collections of medical images [6, 7]. Furthermore, fields with large curated collections have obtained dedicated foundation models, including histopathology [26, 27], ophthalmology [28] , dermatology [29] and remote sensing [30]. Dentistry has followed more recently, with domain-specific models whose pretraining is concentrated in vision due to the scarcity of large dental text corpora [31, 32], and with an earlier contrastive retriever restricted to periapical radiographs [33]. In forensic pathology, a systematic review identified 18 studies of task-specific applications, reporting small samples and variable performance [34], while the work on petrographic images consists of classification and segmentation of thin sections with task-specific models [35, 36]. Therefore, to our knowledge, the available encoders in the three domains offer narrower expertise than the mixed contents of their atlases require, and annotated pairs for training a dedicated one are not available for private collections. The present work examines whether the text already provided by the document can support retrieval when the images are out of domain, under the additional requirements of no training data and of models small enough to run locally, the latter being shared with medical applications that increasingly rely on local open-weight language models [37, 38].

# 3. Methods

## 3.1 Overview

The proposed pipeline receives a collection of PDF books and answers a text query with a ranked list of figures (Fig. 1). During ingestion, each page is parsed into pictures and text blocks, every picture is paired with its caption, and the resulting figures are filtered and indexed through their captions, while at search time the query is embedded and the figures are ranked by their similarity to it. The pairing is the only step involving a generative model, namely the local open-weight Qwen3-VL-8B-Instruct [39], and no page, figure, caption or query is sent to an external service during ingestion or search.

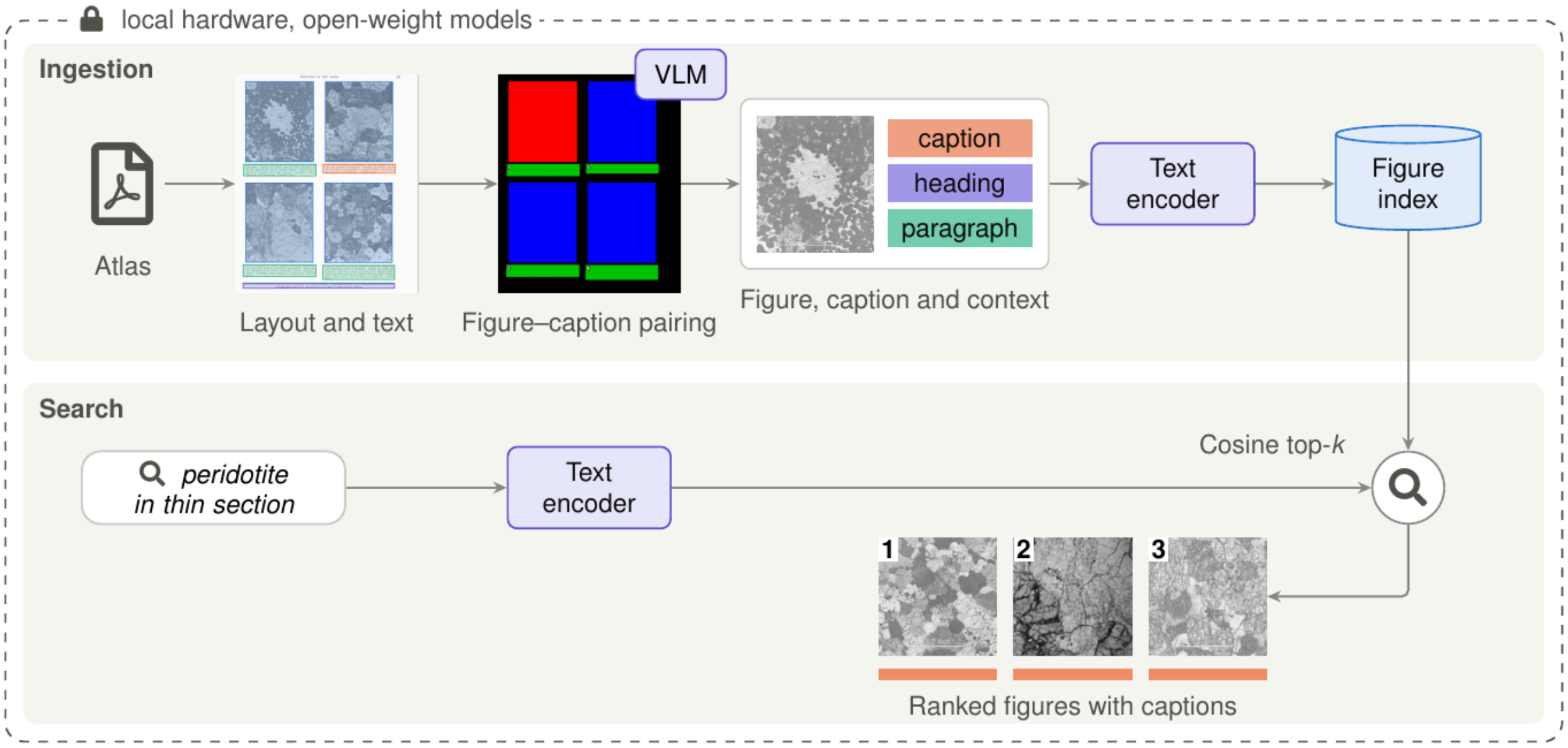

**Fig. 1** Overview of the pipeline. Ingestion parses each page, pairs every picture with its caption from a geometric schematic and indexes the resulting figures. Search embeds the text query and ranks the figures. All steps run on local hardware with open-weight models. Images attributed to [40, 41]

## 3.2 Layout and text

Initially, each page is rendered at 144 DPI, and its regions are located by a YOLOv12-m detector [42] trained on DocLayNet [13]. Of the eleven DocLayNet classes, Picture, Caption and Section-header are used as such, while Text and List-item are treated as body text (Fig. 2). In the case of two largely overlapping picture boxes only the larger is kept, and regions lying inside a picture, such as panel letters and watermarks, are discarded. Regarding the text of every region, it is taken from the PDF text layer, while pages without a usable text layer are read with optical character recognition (OCR) and running headers and footers are removed.

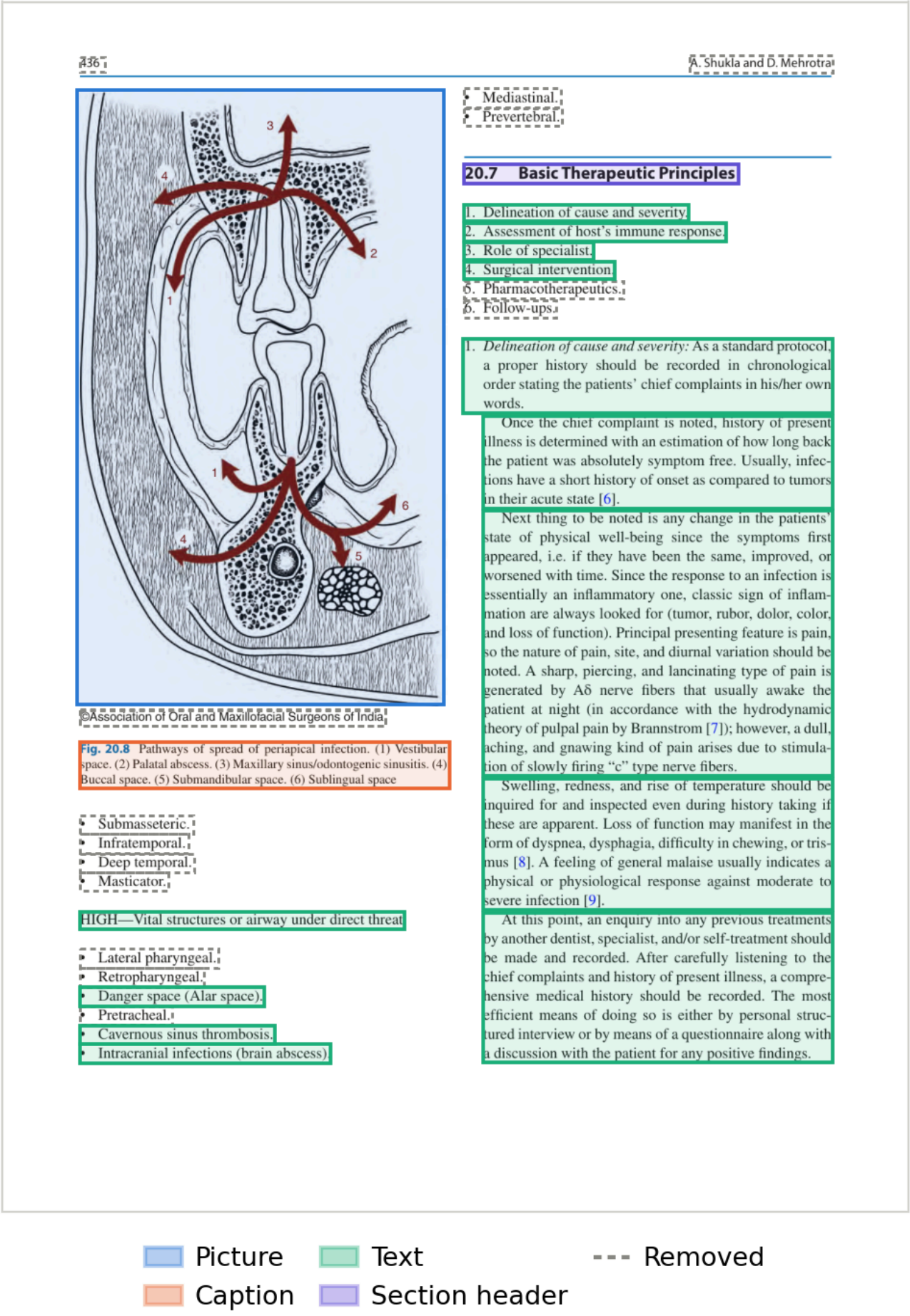

**Fig. 2** Layout parsing of one page. Detected regions are outlined by class, and dashed outlines mark regions removed by the filters. Image attributed to [43]

## 3.3 Figure-caption pairing

Given the figures and text blocks extracted on each page, we pair figures with one item of the candidate caption set, which is comprised of the layout detected caption blocks and any text block lying within 60 px of a picture. Pictures on a page that are paired with the same caption form one figure, each of them being a panel of that figure, while a picture left unpaired becomes a figure without a caption. We filter out icons and decorative strips by requiring figures' short side to be at least 120 px, and an aspect ratio of at most 6.

The VLM is given a schematic of the page in place of the page itself as in Fig. 3. On a black canvas with the dimensions of the page, the target picture is drawn as a red box, the other pictures as blue boxes and each candidate as a numbered green box. The decision therefore rests on the arrangement of the boxes, already fixed by the layout detector, and involves neither the pixels nor the text of the page. This form of pairing is more accurate than the alternatives examined as shown in Section 4.1. Since the schematic contains only box positions, the step can also be assigned to a hosted or closed model without exposing any book content.

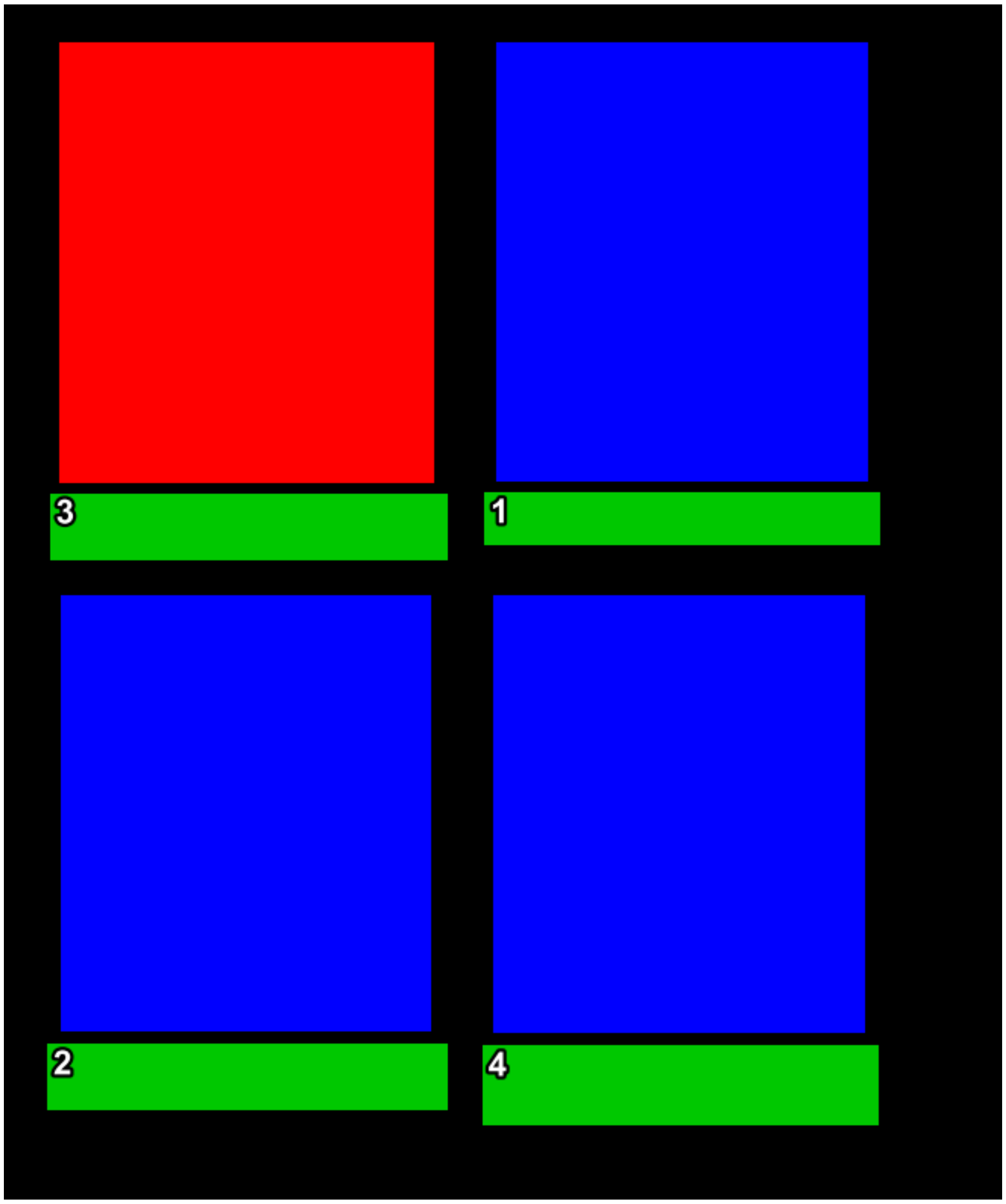

**Fig. 3** Geometric schematic given to the model for one picture. The target picture is red, the other pictures are blue, and the numbered green boxes are the candidate captions

For the evaluation of the geometric pairing, four alternative caption pairing methods are considered. First the same VLM reading of the rendered page on which the pictures and the candidates are outlined, and a heuristic nearest-caption rule. Lastly, the same pages are parsed with Docling 2.151.1 [8] and with MinerU, using its MinerU2.5-Pro model [15, 44], two open document parsers that link pictures to captions after their own layout analysis. The comparison is performed in each domain on ten pages for every topic (Section 3.6.1). We also test two variations of the geometric pairing, the first giving the local VLM the box coordinates as text in place of the drawn schematic, and the second giving the schematic to a hosted model (GPT-5.4-pro [45]).

## 3.4 Indexing and search

Figures are indexed by the caption paired with them (Section 3.3). Captions and queries are embedded with Qwen3-Embedding-0.6B [46], a multilingual text encoder. This caption retrieval route is compared with the other routes in Table 1, including variants that extend the caption with the section heading, with the heading and the nearest paragraph, or with a description of the image written by the local VLM. Regarding the baselines, these include BM25 on the same captions, the generated description alone, and five image-text encoders matching the query

directly to the figure panels, one of them an embedding model built on a VLM, with the two biomedical ones used in the medical domains only. Finally, the caption route is also run with the captions paired by the heuristic rule, by the VLM reading the page and by Docling, and its results are graded on the same scale.

**Table 1** Retrieval routes. Every route answers a text query with a ranked list of figures

| Route | Indexed content | Encoder |
|---|---|---|
| **Proposed** | | |
| Caption | Caption paired by the geometry method | Qwen3-Embedding-0.6B |
| **Variants** | | |
| Caption + heading | Caption and section heading | Qwen3-Embedding-0.6B |
| Caption + heading + paragraph | Caption, section heading and nearest paragraph | Qwen3-Embedding-0.6B |
| Caption + VLM description | Caption and generated description | Qwen3-Embedding-0.6B |
| **Baselines** | | |
| Caption, BM25 | Caption paired by the geometry method | BM25 [47, 48] |
| VLM description | Description written by the local model from the pixels | Qwen3-Embedding-0.6B |
| Pixels, CLIP | Figure panels | CLIP ViT-L/14 [4] |
| Pixels, SigLIP 2 | Figure panels | SigLIP 2 Giant [5] |
| Pixels, Qwen3-VL-Embedding | Figure panels | Qwen3-VL-Embedding-8B [49] |
| Pixels, MedSigLIP | Figure panels (medical domains) | MedSigLIP-448 [7] |
| Pixels, BiomedCLIP | Figure panels (medical domains) | BiomedCLIP [6] |

# 3.5 Data

The pipeline is applied to three collections of reference atlases, in forensic pathology, in dental and oral medicine and in geology (Table 2). The forensic atlases show injuries and post-mortem findings, mostly as photographs of the body and its organs, the dental and oral atlases combine clinical photographs with radiographs and are written in English, Spanish and Portuguese, while the geological ones show rocks and minerals as thin-section micrographs and as photographs of specimens. The books were processed as distributed, with figures repeated in two editions of the same book removed by hand. Fig. 4 showcases the challenging task of retrieving elements for the selected collections.

**Table 2** The three corpora. Kept figures form the candidate set of every route

| | Forensic pathology | Dental and oral medicine | Geology |
|---|---|---|---|
| Books | 10 | 6 | 13 |
| Pages | 3,554 | 1,971 | 3,015 |
| Kept figures | 4,682 | 2,232 | 3,507 |
| Kept figures with a caption | 4,505 (96%) | 2,029 (91%) | 3,416 (97%) |
| Topics | 11 | 16 | 8 |
| Queries | 58 | 80 | 80 |

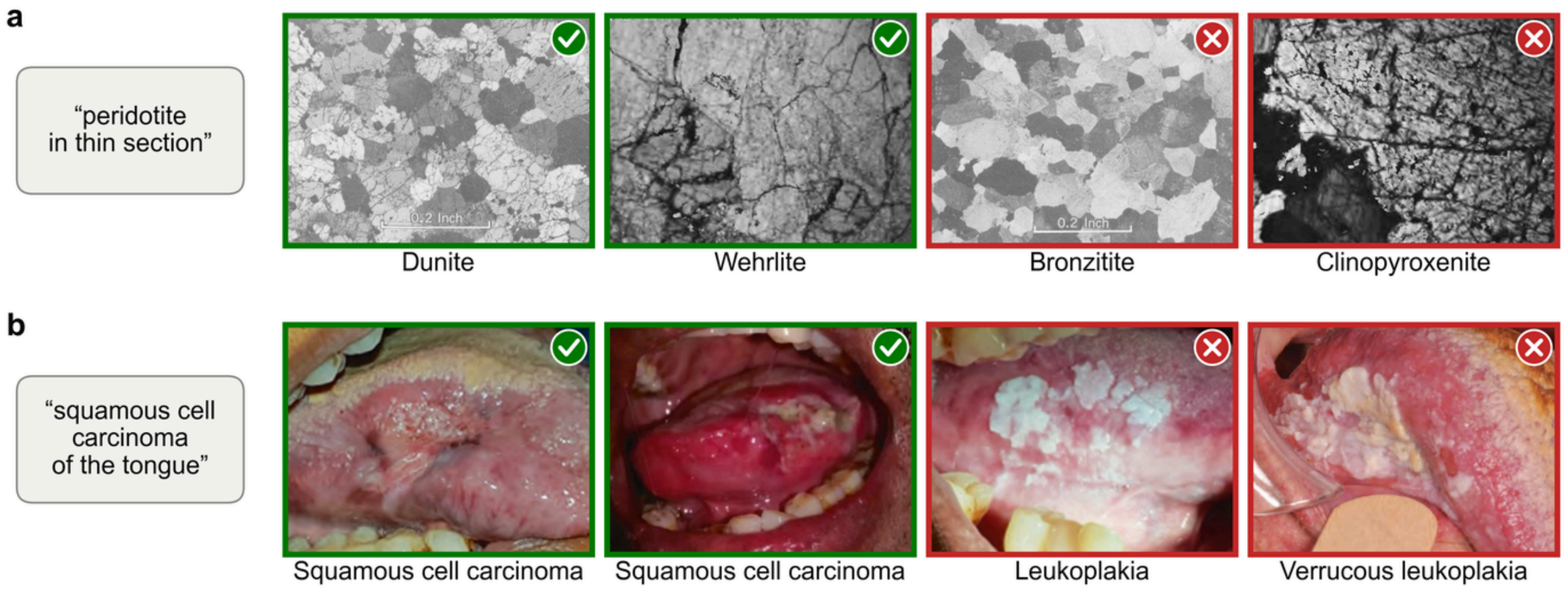

**Fig. 4** Fine-grained distinctions that pixel matching misses. First results of a pixel route for one geology query and one dental query. Green marks a figure that shows the queried finding and red a similar-looking figure that does not. Images attributed to [40, 41, 43]

# 3.6 Evaluation protocol

## 3.6.1 Topics and queries

The chapters of all books in a corpus were manually grouped into topics, and chapters mixing several subjects were left untagged. An established classification was used where feasible, with the dental topics following the chapter scheme of a standard reference [50]. This results in 11 topics in forensic pathology, 16 in dental and oral medicine and 8 in geology. Every figure takes the topic of its chapter, and Fig. 5 shows the distribution of the figures over the topics.

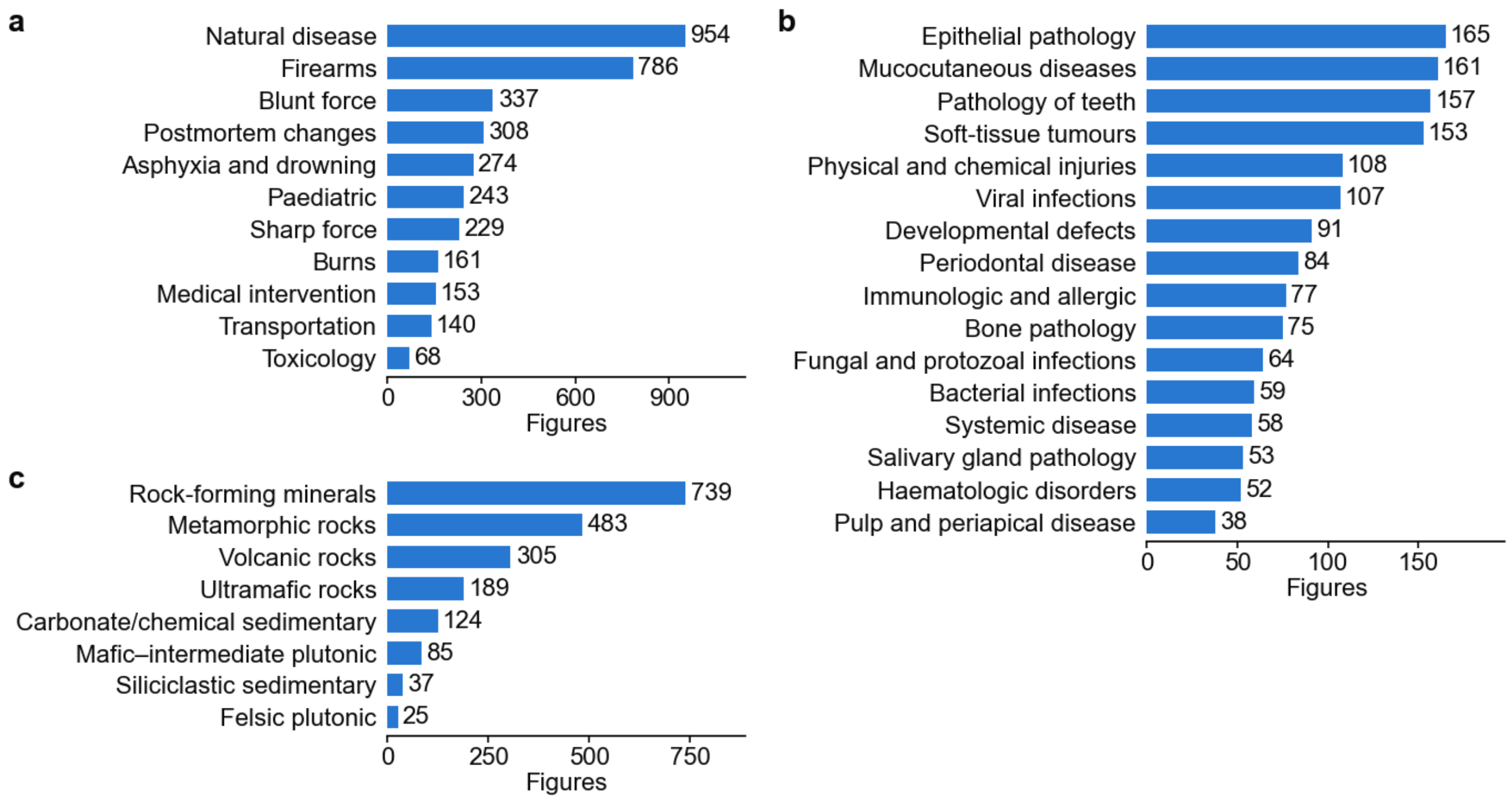

**Fig. 5** Number of kept figures per topic in the three corpora

The retrieval queries used to benchmark the system were drafted by a large language model (GPT-5.4-pro [45]) using only the topic names as context. The model drafted 100 short English queries for each domain naming one finding that an expert might look for, such as "contact gunshot wound of the head with muzzle imprint", "squamous cell carcinoma of the tongue" or "peridotite in thin section". Subsequently, near-duplicate drafts were removed on the query text alone, with no image, caption or retrieval result in view, leaving 58 queries in forensic pathology, 80 in dental and oral medicine and 80 in geology, spread over the topics of each domain. These queries are in English in every domain, providing cross-lingual evaluation in the dental corpus where they meet captions in Spanish and Portuguese. Each query is run against the entire candidate set.

### 3.6.2 Relevance and metrics

For each query, we retrieve the five highest-ranked figures of every retrieval route. For an unbiased grading evaluation, a human grader sees shuffled figures retrieved along with the query, while the retrieval route, the caption, and the item's ranked position are not disclosed. A grade of 0 marks an unrelated figure, 1 a figure on the same topic that does not show the finding asked for, 2 a figure that shows the finding partly or as a secondary element, and 3 a figure that shows exactly the finding asked for. A second reader graded 10% of the pairs again, and the weighted $\kappa$ [51] between the two readers was 0.87 in forensic pathology, 0.77 in dental and oral medicine and 0.84 in geology. The metrics we report are nDCG@5 [52], which takes the grade as gain and the best ordering of the figures graded for the query as its ideal, Hit@1 and Hit@5, where we count queries whose first or top five results, respectively, have a retrieved item with grade of 2 or 3, and the mean grade the average grade of the five results. Finally, alongside the graded scores, a simpler topic score is reported, nDCG@10, with a figure counted as relevant when it belongs to the topic of the query.

## 3.7 Implementation

The code and the prompts of the pipeline are open-source without any corpus content. Model weights are pinned to fixed revisions and loaded from a local cache with network access disabled, and neither ingestion nor search makes an external request. On a consumer-grade GPU, ingestion peaks at about 17 GB of GPU memory, and a text query is answered in 21 ms end-to-end with 1.6 GB of VRAM usage.

# 4. Results

## 4.1 Ingestion

The geometric pairing was correct for 89.9% of the 969 validation pictures, against 84.2% for the VLM reading the rendered page, 72.7% for the heuristic rule, which worked only in forensic pathology, 57.1% for MinerU and 44.2% for Docling (Table 3). Fig. 6 showcases how a distance-based heuristic can fail with example pages where it assigned a neighboring text block instead of the caption, while the VLM's reasoning on the page geometry correctly identifies the caption. Providing the entire page did not help the VLM, which matched the geometric pairing only in dental and oral medicine. The two parsers failed mostly because they left pictures without any caption, with MinerU ahead of Docling in every domain. Geology was the hardest domain for the geometric pairing, which reached 76.8% there. Giving the box coordinates as text instead of the drawn schematic lowered the accuracy of the same VLM to 67.9%, with a loss in every domain. A larger model did not change the result, since the hosted GPT-5.4-pro reached 89.6% with the same schematic.

**Table 3** Figure-caption pairing accuracy. Share of pictures paired with their correct caption (%, Wilson 95% confidence interval) on held-out pages that offer a pairing choice. 10 random pages per topic. The two variations give the local VLM the box coordinates as text, and the schematic to GPT-5.4-pro

| **Method** | **Correct pairing, % (95% CI)** | | | |
|---|---|---|---|---|
| | **FP** | **D/O** | **Geology** | **All** |
| Figures | 319 | 422 | 228 | 969 |
| Geometry | **96.2% [93.5, 97.8]** | **92.2% [89.2, 94.4]** | **76.8% [70.9, 81.8]** | **89.9% [87.8, 91.6]** |
| VLM | 85.3% [81.0, 88.7] | **92.2% [89.2, 94.4]** | 68.0% [61.7, 73.7] | 84.2% [81.8, 86.4] |
| Rule-based | 95.6% [92.8, 97.4] | 60.7% [55.9, 65.2] | 62.7% [56.3, 68.7] | 72.7% [69.8, 75.4] |
| Docling | 68.3% [63.0, 73.2] | 24.6% [20.8, 29.0] | 46.5% [40.1, 53.0] | 44.2% [41.1, 47.3] |
| MinerU | 75.5% [70.5, 79.9] | 45.5% [40.8, 50.3] | 52.6% [46.2, 59.0] | 57.1% [53.9, 60.2] |
| **Variations of the geometric pairing** | | | | |
| Coordinates as text | 79.0% [74.2, 83.1] | 67.8% [63.2, 72.1] | 52.6% [46.2, 59.0] | 67.9% [64.9, 70.8] |
| GPT-5.4-pro | 94.7% | 93.9% | 75.4% | 89.6% [87.2, 91.0] |

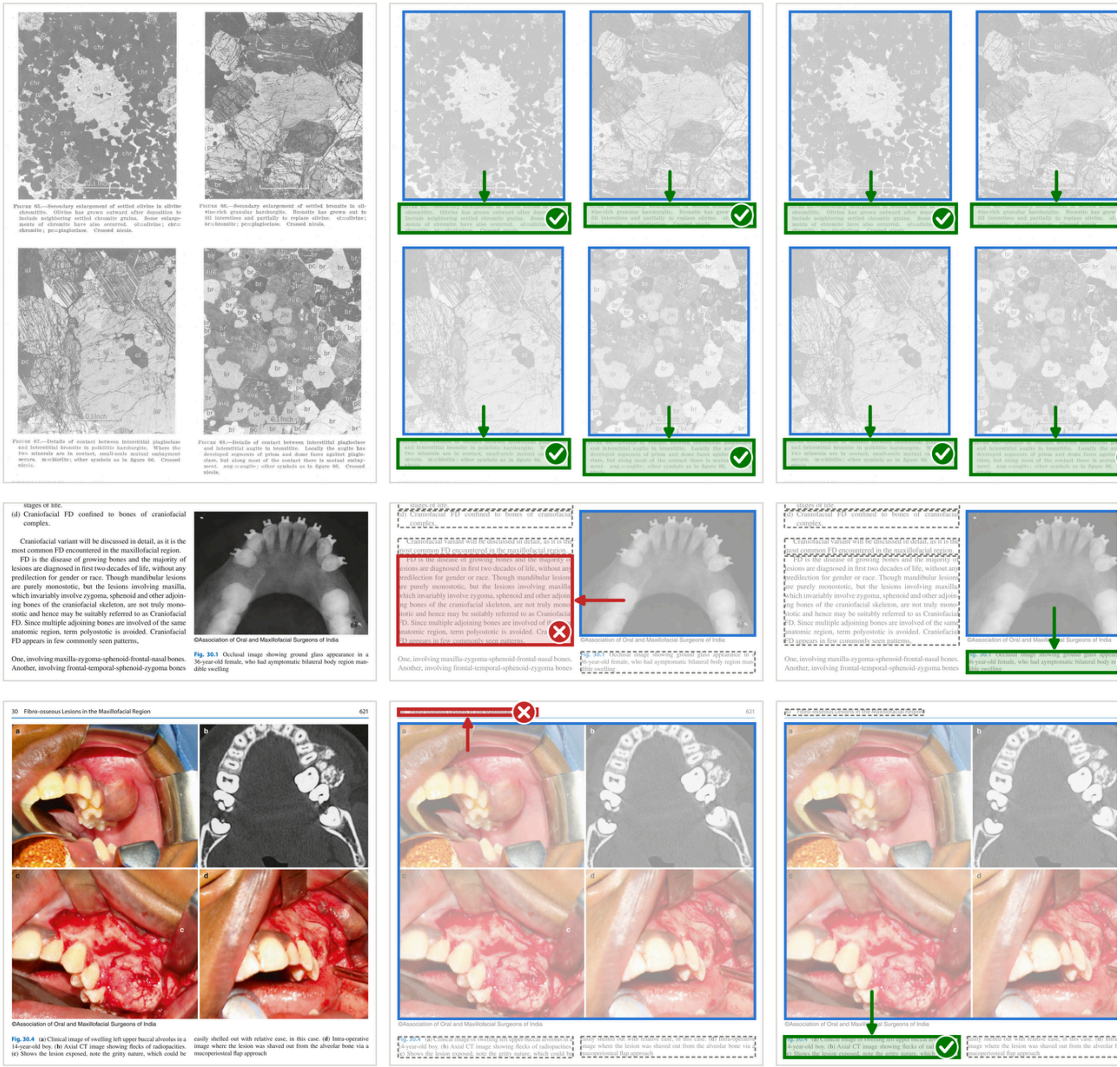

**Fig. 6** Pairing examples. Each row shows a page, the pairing made by the nearest-caption rule and the pairing made by the geometry method. Green marks a correct caption and red a wrong one. Images attributed to [40, 43]

## 4.2 Retrieval

Retrieving figures through their paired captions gave an nDCG@5 of 0.851 in forensic pathology, 0.869 in dental and oral medicine and 0.763 in geology (Table 4). The first figure returned showed the queried finding for more than nine out of ten queries in the two medical domains and for eight out of ten in geology, and a relevant figure was almost always among the first five. BM25 on the same captions (0.766, 0.783 and 0.592) was already better than every route that does not use the captions, and the dense encoder added between 0.085 and 0.171 on top of it.

Matching the query directly to pixels performed much worse, since the best image-text encoder, SigLIP 2, reached 0.540, 0.510 and 0.268 and returned the right finding first for less than half of the queries in every domain. The two biomedical encoders did not help, as both scored below SigLIP 2 in forensic pathology and in dental and oral medicine. Qwen3-VL-Embedding-8B, a larger embedding model built on a VLM, also stayed below SigLIP 2 with 0.467, 0.409 and 0.233, close to it only in geology, and needed about 17 GB of GPU memory at query time. We also tested the descriptions written by the local VLM, which scored 0.413, 0.280 and 0.300 on their own, close to the pixel routes, and did not improve any graded score when appended to the captions.

Adding document context to the caption brought no clear gain on graded relevance, since nDCG@5 was 0.839, 0.853 and 0.783 with the section heading and 0.813, 0.840 and 0.759 with the nearest paragraph as well, in all cases within or below the interval of the caption route. The topic score moved the other way and rose slightly with context (from 0.561, 0.448 and 0.498 to at most 0.583, 0.489 and 0.524), which is expected since the heading and the paragraph describe the subject of the page more than the finding shown in the picture.

Finally, we ran the caption route with the captions paired by three of the other methods, and the differences in pairing accuracy carried over to retrieval. With the geometric pairing the first result showed the queried finding for 91%, 96% and 81% of the queries, compared with 90%, 79% and 66% for the heuristic rule, 86%, 96% and 78% for the VLM reading the page and 81%, 45% and 45% for Docling, which is the same order as in Table 3 in every domain. Docling lost the most, up to 0.39 in nDCG@5 against 0.09 for the rule and 0.06 for the VLM reading the page, because figures left without a caption cannot be retrieved.

**Table 4** Retrieval of figures from text queries in forensic pathology (FP), dental and oral medicine (D/O) and geology. Routes and metrics are defined in Section 3

| Route | Domain | Graded relevance | | | | Topic relevance |
|---|---|---|---|---|---|---|
| | | nDCG@5 [95% CI] | Hit@1 | Hit@5 | Mean grade | Topic nDCG@10 |
| Caption + heading + paragraph | FP | 0.813 [0.774, 0.848] | 0.90 | 0.97 | 2.17 | **0.583** |
| | D/O | 0.840 [0.809, 0.868] | 0.94 | 0.99 | 1.93 | **0.489** |
| | Geology | 0.759 [0.714, 0.800] | 0.78 | 0.95 | 1.86 | 0.516 |
| Caption + heading | FP | 0.839 [0.803, 0.871] | **0.91** | 0.97 | 2.25 | 0.576 |
| | D/O | 0.853 [0.825, 0.878] | 0.93 | **1.00** | 1.97 | 0.484 |
| | Geology | **0.783 [0.744, 0.819]** | 0.78 | **0.99** | **1.91** | **0.524** |
| Caption | FP | **0.851 [0.811, 0.889]** | **0.91** | 0.97 | **2.30** | 0.561 |
| | D/O | **0.869 [0.843, 0.893]** | **0.96** | **1.00** | **1.98** | 0.448 |
| | Geology | 0.763 [0.724, 0.800] | **0.81** | **0.99** | 1.83 | 0.498 |
| Caption, BM25 | FP | 0.766 [0.715, 0.812] | 0.84 | **0.98** | 2.08 | 0.502 |
| | D/O | 0.783 [0.749, 0.817] | 0.88 | 0.97 | 1.77 | 0.347 |
| | Geology | 0.592 [0.532, 0.649] | 0.59 | 0.86 | 1.47 | 0.466 |
| Caption + VLM description | FP | 0.817 [0.772, 0.859] | 0.88 | 0.95 | 2.21 | 0.526 |
| | D/O | 0.848 [0.818, 0.876] | 0.94 | 0.96 | 1.94 | 0.430 |
| | Geology | 0.708 [0.665, 0.749] | 0.76 | 0.95 | 1.69 | 0.481 |
| VLM | FP | 0.413 [0.361, 0.464] | 0.29 | 0.62 | 1.08 | 0.389 |
| description | D/O | 0.280 [0.241, 0.319] | 0.11 | 0.28 | 0.69 | 0.158 |
| | Geology | 0.300 [0.257, 0.342] | 0.21 | 0.42 | 0.75 | 0.276 |
| Pixels, SigLIP 2 | FP | 0.540 [0.481, 0.603] | 0.43 | 0.81 | 1.48 | 0.413 |
| | D/O | 0.510 [0.463, 0.557] | 0.34 | 0.61 | 1.24 | 0.216 |
| | Geology | 0.268 [0.214, 0.323] | 0.16 | 0.38 | 0.64 | 0.254 |
| Pixels, Qwen3-VL-Embedding | FP | 0.467 [0.413, 0.522] | 0.38 | 0.76 | 1.26 | 0.392 |
| | D/O | 0.409 [0.368, 0.448] | 0.20 | 0.50 | 0.99 | 0.155 |
| | Geology | 0.233 [0.190, 0.279] | 0.12 | 0.30 | 0.59 | 0.307 |
| Pixels, CLIP | FP | 0.428 [0.367, 0.489] | 0.41 | 0.69 | 1.13 | 0.351 |
| | D/O | 0.267 [0.226, 0.309] | 0.11 | 0.28 | 0.64 | 0.124 |
| | Geology | 0.146 [0.111, 0.186] | 0.07 | 0.15 | 0.35 | 0.226 |
| Pixels, MedSigLIP | FP | 0.397 [0.334, 0.462] | 0.33 | 0.59 | 1.06 | 0.338 |
| | D/O | 0.359 [0.310, 0.410] | 0.20 | 0.40 | 0.88 | 0.170 |
| Pixels, BiomedCLIP | FP | 0.298 [0.246, 0.357] | 0.21 | 0.50 | 0.79 | 0.235 |
| | D/O | 0.333 [0.291, 0.374] | 0.10 | 0.38 | 0.82 | 0.131 |

# 5. Discussion

## 5.1 Interpretation

Our results show that the caption written by the author is the most useful description of a figure in such collections. Caption retrieval outperformed the best image-text encoder by 0.3 to 0.5 in nDCG@5, and even BM25 on the same captions was better than every route that does not use them, so the gain comes mainly from the caption text and less from the encoder. This agrees with the recent finding that text representations are the most effective in scientific retrieval, even for queries about figures [21]. We extend it to private atlases, where the figure-caption pairs are not available and have to be recovered first.

The image-text encoders could not handle these images, as the best generic encoder stayed below 0.55 in every domain and neither the two biomedical encoders nor a larger embedding model built on a VLM were better. For the biomedical encoders this is consistent with their training on figures from the research literature and on curated clinical images. Even though descriptions generated by a large VLM have been reported to surpass visual document encoders [22], using an 8B model that fits our local constraint did not perform well or improve retrieval when added as an auxiliary signal to the captions. We believe that naming the finding in such pictures requires the knowledge of the specialist who wrote the caption, which a small generic model lacks.

For the pairing, the arrangement of the boxes was enough, since the same VLM was less accurate when reading the rendered page than when seeing only the schematic in two of the three domains, and equally accurate in the third. This suggests that out-of-domain content can distract a small model from a spatial decision. The form of the geometry mattered as well, because the same model lost 22 points when the boxes were listed as coordinates. A much larger hosted model was not more accurate on the schematic, so the remaining errors are probably due to the detected boxes. The heuristic rule could not follow layouts that change from book to book, and the two document parsers linked only 44% and 57% of the pictures to their correct caption, in line with earlier reports of such errors on pages with a non-trivial structure [10].

For a private atlas collection, we therefore recommend indexing each figure by its caption alone. Section headings and neighboring paragraphs moved the results toward the topic of the query but not closer to the finding, so they are useful mainly for browsing figures by subject. The pipeline needs a layout detector, an 8B VLM for the pairing and a 0.6B text encoder, runs on consumer hardware and answers a query in 21 ms. It can be used where the documents cannot leave local hardware and no training data exists, and its pairing step exposes no content even if a hosted model is preferred.

## 5.2 Limitations and future work

The corpora are private, so the reported numbers cannot be reproduced on the same data. The released code and prompts allow the pipeline to be applied to other collections. Three domains and 29 books do not cover every atlas layout, and the lower pairing accuracy in geology shows that the method depends on the layout detector and on the caption conventions of each book. Figures without a caption, between 3% and 9% of the kept figures, cannot be retrieved by the caption route.

Regarding the evaluation, relevance was graded by a single grader, with a second reading of 10% of the pairs, and only for the first five results of each route, so a new route would require new grades. The queries were drafted by a language model, are written in English and name a finding in the vocabulary of the field. Such queries favor routes that match text, and queries phrased by non-specialists or given as images may reduce the distance to the pixel routes. The topic score is coarse because it is defined at the level of chapters. Finally, only one of the five image-text encoders is an embedding model built on a VLM.

In future work, we will address image queries and the combination of routes, where generated descriptions or pixels could serve the figures without a caption.

# 6. Conclusion

This work presented a structure-aware pipeline for retrieving figures from expert atlases, using only locally run open-weight models and no in-domain training. The pipeline recovers the caption of each figure from the structure of the page and retrieves figures through these captions. To pair figures with captions we introduced a geometric method, where an 8B VLM sees only a schematic of the detected boxes and no page content. This method was correct for 89.9% of the validation pictures, more accurate than the same VLM reading the page and a heuristic rule, and well above the two document parsers, MinerU (57.1%) and Docling (44.2%). A hosted model given the same schematic was equally accurate, so this step can be delegated without exposing content. In the three domains we examined, caption retrieval reached an nDCG@5 between 0.76 and 0.87, while the best image-text encoder stayed between 0.27 and 0.54 and the descriptions generated by the same VLM did not exceed 0.42. We conclude that for private collections without labels or a suitable image encoder, recovering the captions from the page structure is a more reliable basis for retrieval than matching the pixels. The code and the prompts are open-source.